\documentclass[conference,compsoc]{IEEEtran}
\usepackage[all]{pdfprivacy}

\ifCLASSOPTIONcompsoc
  \usepackage[nocompress]{cite}
\else
  \usepackage{cite}
\fi

\usepackage{graphicx}             
\usepackage{amssymb}              
\usepackage{amsmath}              
\usepackage{xspace}               
\usepackage{booktabs}             
\usepackage{microtype}            
\usepackage{paralist}             
\usepackage{pifont}               
\usepackage{xurl}                 
\usepackage[hidelinks]{hyperref}  

\newcommand{\cmark}{\ding{51}}
\newcommand{\xmark}{\ding{55}}

\newcommand{\ourMethodNoSpace}{WoE}
\newcommand{\ourMethod}{\ourMethodNoSpace\xspace}

\makeatletter
\newcommand{\linebreakand}{%
  \end{@IEEEauthorhalign}
  \hfill\mbox{}\par
  \mbox{}\hfill\begin{@IEEEauthorhalign}
}
\makeatother

\begin{document}

\title{WoE Wrote It? Watermarking Mixture-of-Experts LLMs\\ for Black-Box Text Provenance}



\author{
    \IEEEauthorblockN{Jona te Lintelo}
    \IEEEauthorblockA{\textit{Radboud University}}
    \and
    \IEEEauthorblockN{Lichao Wu}
    \IEEEauthorblockA{\textit{University of Bristol}}
    \linebreakand
    \IEEEauthorblockN{Stjepan Picek}
    \IEEEauthorblockA{\textit{Faculty of Electrical Engineering and Computing, University of Zagreb} \& \\
    \textit{Radboud University}}
}

    

\maketitle

\begin{abstract}
Large Language Model (LLM) watermarks provide a mechanism for text provenance, enabling model owners to identify machine-generated content and attribute it to a specific watermarked model. However, current LLM watermarking approaches predominantly rely on inference-time sampler methods and focus their analysis on dense models. Inference-time methods are only effective when the text is explicitly generated via the model owner's controlled API; they fail in a post-theft scenario. An adversary who steals or leaks the model weights gains complete control over inference and can simply run an unmodified sampler, bypassing the watermark and preventing post-theft attribution. In this work, we introduce Watermarking of Experts (WoE), a novel black-box text provenance method that leverages the unique structural properties of sparse Mixture-of-Experts (MoE) models. WoE biases the vocabulary of a few secret experts, and defenders detect it by replaying suspect text through their own checkpoint to recover where these experts acted. This approach ensures the watermark remains intrinsic to the model parameters, enabling defenders to attribute text generated by stolen weights, leaked checkpoints, and secondary dense models distilled from the stolen model without needing access to the adversary's deployment or weights. We evaluate WoE across eight MoE models, demonstrating successful watermark detection from suspect text, achieving an average true positive rate of 92.5\% at a 1\% false positive rate, reaching up to 95.3\%, while reducing benchmark accuracy by only 0.4\% on average. Unlike prior MoE watermarks, WoE survives distillation into a dense student model and remains detectable under adversarial fine-tuning, forcing malicious actors into a trade-off in which weakening the attribution signal requires additional model adaptation or compromises model utility.
\end{abstract}

\section{Introduction}
\label{sec:introduction}
Mixture-of-Experts (MoE) architectures have become increasingly prevalent in state-of-the-art LLMs, enabling greater model capacity through sparse expert activation at lower computational cost~\cite{shazeer2017moe,fedus2022switch}. As these models grow in capability and value, their weights become attractive targets for intellectual property theft~\cite{li2023protectingip}. Once an adversary obtains the weights, they can independently deploy, modify, or fine-tune the model, bypassing mechanisms controlled by the original model owner. This creates a critical challenge: establishing \emph{post-theft text provenance}, i.e., determining whether suspect text is statistically attributable to a specific stolen model or its derivatives, solely from its outputs and without access to the adversary's deployment.

\textbf{Text watermarks and their pitfalls.} Text watermarking embeds statistically detectable signals into generated text to establish model provenance~\cite{kirchenbauer2023watermark,gloaguen2026unifiedframeworkllmwatermarks}. Major AI providers have already deployed or announced watermarking mechanisms~\cite{google2024synthid,anthropic2026watermark}, while large-scale model extraction has been documented in practice~\cite{anthropic2026distillation}. Despite their growing practical importance, existing watermarking approaches face fundamental limitations in post-theft scenarios. First, \emph{inference-time watermarks} modify the decoding process to introduce detectable biases into generated text~\cite{kirchenbauer2023watermark,he2025theoretically}. However, these mechanisms rely on control over the inference pipeline, which is lost once an adversary obtains the model weights. By deploying the stolen model with an unmodified sampler, the adversary can bypass the watermark entirely. Second, \emph{weight-embedded watermarks} incorporate detectable signals directly into model parameters~\cite{block2025gaussmark,li2023quantizewatermarking}, allowing them to persist independently of the decoding process. However, existing methods designed for dense architectures typically modify components activated for every token, causing the watermark to influence the model's output distribution globally. This creates a fundamental trade-off: stronger watermark signals improve detectability but can distort generation quality and degrade downstream model utility. Moreover, broadly distributed watermark effects may be weakened through subsequent fine-tuning, allowing adversaries to suppress attribution while retaining useful model capabilities (Sections~\ref{subsec:experiments_comparison} and~\ref{subsec:fine_tuning}). These limitations motivate a different design principle: rather than continuously perturbing the model's predictions, a watermark should be \emph{embedded in model weights but activated selectively}, reducing its impact on normal generation while retaining a statistically detectable signal. Crucially, the legitimate owner must be able to recover this signal from suspect text alone, without accessing or querying the adversary's model. Achieving such \emph{zero-access, black-box text provenance} requires an architecture-aware approach that reconciles watermark detectability, robustness, and model utility.

\textbf{Our solution.} We introduce \emph{Watermarking of Experts} (\ourMethod), a weight-embedded watermarking framework that enables model-specific, black-box text provenance for MoE LLMs, even after model theft. \ourMethod exploits two complementary properties of MoE architectures: \emph{sparse expert activation} allows a watermark to influence predictions selectively, while \emph{deterministic expert routing} enables the defender to reconstruct where the watermark was activated from suspect text. Specifically, \ourMethod embeds a secret vocabulary bias into a small number of selected experts, such that the watermark primarily influences token predictions when these experts are activated. Unlike conventional weight-embedded approaches that affect every token, this selective mechanism limits the watermark's impact on model utility. Moreover, since the watermark resides directly in expert weights, it persists independently of the inference pipeline and requires no additional computation during generation.

The \ourMethod pipeline consists of three stages. First, \emph{payload creation} derives a pseudorandom green list and selects target experts using a defender-held secret seed. Second, \emph{watermark injection} uses Low-Rank Adaptation (LoRA)~\cite{hu2022lora} to train only the selected experts to increase the probability of green-list tokens while preserving the model's original behavior elsewhere. Third, \emph{watermark detection} replays suspect text through the defender's watermarked checkpoint to reconstruct expert routing and identify positions where the watermark may have influenced generation. It then measures whether the subsequent tokens exhibit a statistically significant green-list bias relative to the clean model's predictions, using a calibrated detection statistic and a threshold established from diverse innocent-text sources. Consequently, the defender can attribute suspect text to the watermarked model using only the text, without querying or accessing the adversary's deployment or model weights. Our main contributions are:
\begin{compactitem}
  \item We propose \ourMethodNoSpace, an architecture-aware watermarking framework that embeds a secret vocabulary bias into selected MoE experts through localized parameter adaptation. By exploiting sparse expert activation, the watermark remains embedded in model weights while selectively influencing token predictions, leaving a statistical signal that the legitimate model owner can recover from generated text.
  \item We develop a routing-reconstruction-based detector that recovers watermark activation positions from suspect text and identifies statistical deviations from the clean model's predictions. Combined with calibrated thresholds across diverse innocent-text sources, it enables post-theft attribution without access to the adversary's model or inference pipeline.
  \item Across eight MoE models, \ourMethod achieves an average TPR of 92.5\% at a nominal 1\% FPR, with only a 0.4\% accuracy reduction and a 14.0\% perplexity increase. It outperforms existing weight-embedded baselines in detection and utility preservation, supported by extensive ablation studies.
  \item We evaluate \ourMethod against six categories of post-theft attacks, including expert masking, target localization, fine-tuning, quantization, distillation, and paraphrasing. \ourMethod retains 60.4--65.5\% TPR after 1,000 fine-tuning steps and 79.2--81.1\% TPR after dense-model distillation, demonstrating robustness and revealing the costs and limitations of watermark removal.
\end{compactitem}

The code to perform \ourMethod is publicly available at~\url{https://github.com/jonatelintelo/WatermarkingOfExperts}.

\section{Background}
\label{sec:background}

\subsection{LLM Text Watermarking}
\label{subsec:llm_text_watermaking}

Existing LLM text watermarking methods differ in where the watermark is embedded, how it influences generation, and what information is required for detection. Table~\ref{tab:related_work} summarizes these distinctions, particularly their implications for post-theft text provenance. \ourMethod is the only approach that combines selective watermark activation, text-only detection, and persistence after model theft and distillation.

\textbf{Inference-time watermarking} modifies the sampling procedure to bias the next-token distribution during generation~\cite{kirchenbauer2023watermark,gloaguen2026unifiedframeworkllmwatermarks,he2025theoretically}. A representative approach, KGW~\cite{kirchenbauer2023watermark}, pseudorandomly partitions the vocabulary into green and red lists based on preceding tokens and increases the logits of green-list tokens by a fixed bias $\delta$. Detection reconstructs these partitions from suspect text and tests whether green-list tokens occur more frequently than expected. However, such methods depend on the inference pipeline and can be bypassed when an adversary controls the sampler.

\textbf{Weight-embedded watermarking} incorporates the watermark directly into model parameters, allowing it to persist independently of the sampling procedure. Distillation-based approaches train a student model on watermarked teacher outputs to transfer the watermark into its weights~\cite{gu2024learnability,gloaguen2025opensource}. Other approaches directly modify model parameters through weight quantization or perturbations~\cite{li2023quantizewatermarking,block2025gaussmark}. Unlike inference-time methods, weight-embedded watermarks can survive model theft, but existing dense-model approaches typically influence every token, creating trade-offs between detectability, model utility, and robustness.

\begin{table*}[!t]
  \centering
  \footnotesize
  \caption{Comparison of LLM watermarking methods for post-theft text provenance.}
  \label{tab:related_work}
    \begin{tabular}{llllcc}
    \toprule
    Method & Signal lives in & Active on & Defender needs & Survives weight theft & Survives distillation \\
    \midrule
    KGW~\cite{kirchenbauer2023watermark} & Sampler & Every token & Published text & \xmark & \xmark \\
    KGW-D~\cite{gu2024learnability} & All weights & Every token & Published text & \cmark & \cmark \\
    GaussMark~\cite{block2025gaussmark} & One weight matrix & Every token & Published text & \cmark & \cmark \\
    PathMark~\cite{gao2026pathmark} & Routing on trigger inputs & Trigger inputs only & Queries or weights & \cmark & \xmark \\
    WaterMoE~\cite{sun2026watermoe} & Router bias at inference & Every token & Published text & \xmark & \xmark \\
    \midrule
    \textbf{WoE (ours)} & Secret experts & Tokens routed to the experts & Published text & \cmark & \cmark \\
    \bottomrule
  \end{tabular}
\end{table*}

\subsection{Mixture-of-Experts Architectures}
\label{subsec:moe_architectures}

Conventional LLMs rely on dense architectures that activate all parameters for every input token, incurring high computational costs. To address this scaling bottleneck, MoE architectures introduce conditional computation. MoE architectures replace the single dense feed-forward network (FFN) in each Transformer block with a set of independent sub-networks called experts~\cite{shazeer2017moe,fedus2022switch}.
A core component of the MoE architecture is the router, which dynamically dictates which experts process a specific token. For each input token, the router computes a set of unnormalized routing logits over the available experts. To ensure sparsity, the router uses a top-$k$ selection mechanism, identifying and activating only the $k$ experts with the highest routing scores. The remaining experts stay inactive, allowing total parameter capacity to expand while maintaining a relatively small, fixed computational budget per forward pass. The final output of the MoE layer is then an aggregate of the outputs from these selected experts.
Recent MoE designs add structural variations to further increase performance while maintaining sparsity. The main distinction is between standard sparse MoE and shared-expert MoE architectures. In a standard sparse MoE architecture, all available experts are subject to the router's dynamic top-$k$ selection. Shared-expert architectures contain a limited number of experts that bypass the top-$k$ routing mechanism entirely and remain active for all tokens. This continuous activation helps capture common, general-purpose linguistic knowledge and reduces knowledge redundancy among the sparse experts.

\section{Threat Model}
\label{sec:threat_model}

We consider a \emph{post-theft scenario} in which an adversary obtains the weights of a watermarked MoE model and generates text outside the model owner's control, either directly using the stolen model or through a distilled derivative. The defender aims to determine whether suspect text is statistically attributable to the watermarked model, without accessing the adversary's deployment.

\textbf{Attacker.} We assume a strong adversary with full white-box access to the watermarked model and knowledge of \ourMethod, including its candidate layers and expert selection strategy. However, the adversary does not possess the original clean checkpoint or the defender's secret seed $K$, which determines both the green-list vocabulary $G$ and the target expert configuration $C$ (Section~\ref{subsec:payload_creation}). The adversary can freely control inference, inspect and modify model weights, fine-tune or distill the model, and transform generated text. We consider adaptive attacks targeting the watermark, including expert masking, target localization, and watermark removal through fine-tuning (Section~\ref{sec:security_evaluation}).

\textbf{Defender.} We assume a \emph{zero-access, black-box} detection setting. The defender observes only suspect text and cannot access or query the adversary's model or inference pipeline. The defender retains the clean and watermarked checkpoints and the secret seed $K$, enabling reconstruction and statistical verification of the embedded watermark. The watermark is merged into the model weights before deployment, preventing trivial removal by deleting separate adapters.

\section{Structural MoE Watermarking}
\label{sec:methodology}

A watermark that survives weight theft must live in the weights, and a watermark that a zero-access defender can verify must be readable from text alone. MoE architectures make it possible to meet both requirements with a small and secret footprint. 
Specifically, \ourMethod leverages the sparse routing pathways of the MoE architecture and embeds the watermark in expert weights. Every token passes through only a few experts per MoE layer, so a watermark trained into a handful of experts changes only their weights, and its effect enters the computation only where the router selects one of them. The routing that decides where the watermark enters also lets the defender find it again. As routing is deterministic, replaying a suspect text through the watermarked checkpoint reveals where the target experts acted. 

\begin{figure*}[!t]
  \centering
  \includegraphics{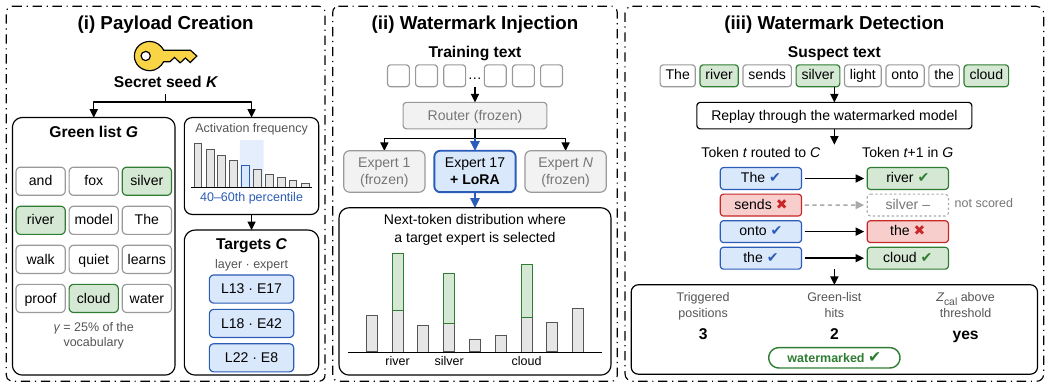}
  \caption{Overview of \ourMethodNoSpace. (i) \emph{Payload creation} derives a green list $G$ and target expert configuration $C$ from a secret seed $K$. (ii) \emph{Watermark injection} uses LoRA to embed a green-list bias into selected experts, activated through sparse routing. (iii) \emph{Watermark detection} reconstructs the union mask $U$ from suspect text and compares the resulting statistic $Z_{\mathrm{cal}}$ against a calibrated threshold.}
  \label{fig:method_figure}
\end{figure*}

Turning this observation into a watermark raises three problems: (i) the adversary must not learn which experts carry the signal; (ii) a few experts must learn to shift the next-token distribution; and (iii) the detector must separate watermarked text from every kind of innocent text at a controlled FPR. \ourMethod addresses them in three stages, shown in Figure~\ref{fig:method_figure}. First, \emph{payload creation} derives a green list and a secret target configuration of experts from the defender's seed (Section~\ref{subsec:payload_creation}). Next, \emph{watermark injection} trains the target experts to bias them towards producing certain tokens (Section~\ref{subsec:watermark_injection}). Finally, \emph{watermark detection} replays the routing of suspect text, scores the tokens that follow the triggered positions against what the clean model would have produced (Section~\ref{subsec:watermark_detection}), and compares the score with a threshold calibrated on four sources of innocent text (Section~\ref{subsec:thresholds}).

\subsection{Payload Creation}
\label{subsec:payload_creation}

The foundation of \ourMethod is the payload $P$, which comprises a pseudorandom green list $G$ and a sparse target configuration $C$ of layer--expert pairs, both derived from the defender's secret seed $K$. 

\textbf{Green List.} The first component of the payload is the green list, which defines the vocabulary bias that the watermark will carry. Let $\mathcal{V}$ denote the vocabulary of the tokenizer and $\mathcal{S}\subset\mathcal{V}$ the special tokens (e.g., padding, EOS, BOS) in that vocabulary. To ensure the watermark does not interfere with structural tokens during generation, we use $K$ to sample the green list without special tokens, $G\subset\mathcal{V}\setminus\mathcal{S}$, uniformly at random such that $|G|=\lfloor\gamma|\mathcal{V}\setminus\mathcal{S}|\rfloor$, for a green-list fraction $\gamma$. Excluding the special tokens keeps the watermark away from tokens that control the structure of the output; a green end-of-sequence token, for example, would let the watermark decide where texts end. Drawing the list uniformly at random, independently of the model and of any text, makes the membership of each token independent of its frequency and meaning. The green-list test rests on exactly this, since its null hypothesis is text written without knowledge of the list~\cite{kirchenbauer2023watermark}. The fraction $\gamma$ trades evidence against flexibility. At a fixed bias, a smaller list yields more evidence per token but leaves fewer green alternatives at positions where the model is nearly certain~\cite{kirchenbauer2023watermark}.

\textbf{Target Configuration.} With the green list $G$ defined, the next challenge is to identify the target layers and experts in which to embed the vocabulary bias. The configuration must be sparse, so the watermark enters the computation only where routing passes through a few experts. It also has to be secret, as otherwise an adversary who knows the target experts could remove or retrain them while leaving the rest of the model intact.
To resolve this, our framework draws the configuration with the secret seed $K$ within candidate ranges for both layers and experts, so that many configurations are possible. On Qwen3.6-35B-A3B, for example, the three target experts lie hidden among 1,020 candidate experts (Table~\ref{tab:model_specifications} lists the candidates of every model). An adversary who masks the three experts of a guessed configuration cannot tell whether it hit the targets, because testing for the watermark requires the green list, which the seed also hides. An adversary who knows the algorithm therefore knows the possible target configurations, but not which of them $K$ selected, and its blind alternative is to mask all candidate experts at once; Section~\ref{subsec:broad_masking} evaluates this attack.

We take the candidate layers from the second half of the MoE layers, because mechanistic interpretability research shows that late layers are primarily responsible for the final vocabulary projection~\cite{geva2022transformer,lanquillon2026gptrefusal}. A target expert has to raise the probability of green tokens among the tokens the model already considers likely, which is a correction of the next-token distribution itself, so it is best placed where that distribution is formed. An expert in an early layer would instead have to plant a change that survives every later layer. In our preliminary experiments, experts in early and middle layers learned no usable watermark at any rank, learning rate, or training length we tried, whereas experts in late layers did, and our depth ablation confirms this at the default settings (Section~\ref{subsec:ablation_layer}). Let $L_{\mathrm{target}}$ denote the resulting set of target layers drawn from the second half using the secret seed $K$. 

Having selected $L_{\mathrm{target}}$, the next step is to determine which experts, and how many, to target within each layer to balance watermark detectability and model utility. Since modern MoE architectures contain many experts, an exhaustive search is infeasible as one target expert in each of the three target layers gives $N^3$ configurations for $N$ routed experts per layer ($2.6\times10^{5}$ for OLMoE-1B-7B-0125-Instruct, $1.7\times10^{7}$ for Qwen3.6-35B-A3B), each requiring its own injection run. We instead derive both choices from how routing behaves. Concretely, the decision of which experts to target follows from their activation frequency, which governs both how often the watermark acts and how well training can inject it. A rarely activated expert is selected for few tokens, so training sees few positions to learn the vocabulary bias from, and the weaker bias it learns acts on few tokens of a text. A frequently activated ``core’’ expert is selected for a broad range of tokens, so its low-rank update must serve inputs too varied to push each of them toward the green list, and the bias takes hold only partly. We therefore draw the target expert from the middle band of the activation-frequency ranking, where an expert is selected often enough to be trained and for inputs specific enough for the bias to take hold. Section~\ref{subsec:ablation_expert} confirms that targets from either end carry a far weaker watermark.

Next, the decision of how many experts to target follows from the top-$k$ routing mechanism and its distribution. In MoE models, the top-$k$ routing mechanism selects multiple experts for the same token, and expert activation is shown to be imbalanced, with a few experts being selected far more often than the others~\cite{wu2026gatebreaker}. Every additional target expert in a layer adds the tokens routed to it, so the vocabulary bias reaches a larger share of the generated text and costs more generation quality. The detector gains less from these experts. One target expert per layer already triggers often enough for a reliable test, and top-$k$ routing can select several target experts for the same token, so that their router-weighted contributions overlap at the same prediction position instead of adding positions for the detector. We therefore target one expert per layer, which keeps the watermark sparse, and Section~\ref{subsec:ablation_expert} measures two and eight experts per layer.

Let $e^*_l$ be the target expert $e^*$ in layer $l$. We construct the target configuration $C$ and the resulting payload $P$ as
\begin{equation}\label{eq:payload}
P=(G,\,C),\qquad C=\{(l,e_l^*):l\in L_{\mathrm{target}}\}.
\end{equation}
This payload is used twice. Watermark injection trains the target experts in $C$ to shift the next-token distribution toward $G$ (Section~\ref{subsec:watermark_injection}), and watermark detection uses $C$ to find the positions at which the target experts acted and $G$ to score the tokens that follow them (Section~\ref{subsec:watermark_detection}).

\subsection{Watermark Injection}
\label{subsec:watermark_injection}

Once the payload is established, the target experts must learn to bias the text toward $G$, but only where the router selects them. A decoding-time watermark can add the vocabulary bias to the logits directly, whereas a target expert cannot. Its output is scaled by its routing weight, combined with the outputs of the other selected experts, and added to the residual stream, which every later layer reads and modifies before the vocabulary projection. At each position where it is selected, a target expert therefore has to learn an output that still shifts the next-token distribution toward $G$ after passing through all later layers.

\textbf{Triggered Positions.} To apply the watermark only at these positions, the training objective first has to identify them. For the token at position $t$, let $R_{l,t}$ denote the set of experts that the router of layer $l$ selects, the $k$ experts with the highest routing scores~\cite{shazeer2017moe,fedus2022switch}. We then construct a binary \emph{union mask} $U$ that indicates whether the token passes through any target expert, defined as
\begin{equation}\label{eq:union_mask}
U_t=\bigvee_{(l,e)\in C}\mathbb{I}[e\in R_{l,t}].
\end{equation}
A token does not need to activate all target experts; activating any one of them sets $U_t=1$. Processing token $x_t$ produces the logits used to predict $x_{t+1}$, so the triggered prediction positions are
\begin{equation}\label{eq:triggered_positions}
\mathcal{I}=\{t:U_t=1\},\qquad N_{\mathrm{trig}}=|\mathcal{I}|.
\end{equation}
The share of triggered positions in a text is its \emph{trigger rate}.

\textbf{Injection Objective.} \ourMethod embeds the payload by fine-tuning the target experts through LoRA adapters on the gate, up, and down projections of their feed-forward networks. All other components remain frozen, including the routers, the attention layers, the other experts, and the vocabulary projection. This architectural isolation localizes the immediate watermark bias to positions routed through a target expert. Freezing them prevents globally active components from learning the watermark; altering globally active parameters directly degrades general model utility (Section~\ref{subsec:experiments_comparison}).

The next step is deciding what the target experts should learn at a triggered position. A decoding-time KGW watermark~\cite{kirchenbauer2023watermark} with green list $G$ adds $\delta$ to the logits $z_t$ of the green tokens and turns the clean next-token distribution $p_t=\mathrm{softmax}(z_t)$ into
\begin{equation}\label{eq:kgw_target}
q^*_t=\mathrm{softmax}(z_t+\delta g),
\end{equation}
where $g\in\{0,1\}^{|\mathcal{V}|}$ indicates the green tokens and $\delta$ sets the watermark strength. This distribution solves $\max_q \mathbb{E}_{v\sim q}[g_v]-\frac{1}{\delta}\mathrm{KL}(q\,\|\,p_t)$, which is the optimal trade-off between green mass and distortion of the clean distribution~\cite{gloaguen2026unifiedframeworkllmwatermarks}. 

\ourMethod distills $q^*_t$ into the target experts. The model is trained toward $q^*_t$ at triggered positions and toward the clean distribution at all other positions, with the objective
\begin{equation}\label{eq:injection_loss}
\mathcal{L}=\frac{1}{N_{\mathrm{pred}}}\Big[\sum_{t\notin\mathcal{I}}\mathrm{KL}\big(p_t\,\|\,p_{\theta,t}\big)+\sum_{t\in\mathcal{I}}\mathrm{KL}\big(q^*_t\,\|\,p_{\theta,t}\big)\Big],
\end{equation}
where $p_{\theta,t}$ is the next-token distribution of the model being trained, $t$ runs over the $N_{\mathrm{pred}}$ valid next-token prediction positions of a batch, $\mathcal{I}$ is read from the model's own routing in the same forward pass, and $p_t$ comes from the same model with the adapters disabled. The first term anchors the model to the clean distribution wherever the watermark should be absent; it is a training objective, not a guarantee that these positions remain unchanged. The second term teaches the target experts the distribution that a KGW sampler would produce at the positions they touch. After training, the adapters are merged into the expert weights.

\subsection{Watermark Detection}
\label{subsec:watermark_detection}

To detect the watermark without access to the adversary's deployment, API, or weights, we exploit the deterministic nature of causal Transformer routing. When a suspect text sequence is observed, the defender processes it with the watermarked checkpoint to reconstruct its internal routing and find the positions at which the target experts were selected.
Indeed, the router's choice at position $t$ depends only on the tokens up to $t$. Replaying the same text through an unmodified copy of the watermarked checkpoint therefore reproduces the routing that produced the text, and with it the union mask $U$ (Equation~\ref{eq:union_mask}) and the triggered positions $\mathcal{I}$. The detector scores only positions whose next token $x_{t+1}$ belongs to the suspect text, since only those tokens were produced by the suspect model. Small numerical differences between generation and replay can flip routing decisions between nearly tied experts; the detection rates we report include this effect, because every watermarked text in our evaluation is generated and then replayed. If the adversary has modified the weights, its routing may differ further from the replay, and Section~\ref{sec:security_evaluation} examines such modifications.

\textbf{Calibrated Detection Statistic.} With the triggered positions $\mathcal{I}$ reconstructed, the detector tests at each of them whether the next observed token $x_{t+1}$ belongs to the green list. The number of green-list hits is
\begin{equation}\label{eq:hit_count}
H=\sum_{t\in\mathcal{I}}\mathbb{I}[x_{t+1}\in G].
\end{equation}
To judge whether these hits are evidence of the watermark, we compare them with the null hypothesis that the text is unwatermarked (e.g., written by a human or generated by another model). Under this null hypothesis, the number of hits to expect is not the same for every suspect text. Because the green list is fixed, a text whose most likely next tokens are in the green list collects many hits without any watermark. \ourMethod therefore measures $H$ against the clean model, which the defender owns. Let $\mu_t=\sum_{v\in G}p_t(v)$ be the clean model's green mass at position $t$. If $x_{t+1}$ were drawn from the clean model, the hit at $t$ would be a Bernoulli variable with mean $\mu_t$ and variance $\mu_t(1-\mu_t)$, which gives the calibrated detection statistic
\begin{equation}\label{eq:detection_score}
Z_{\mathrm{cal}}=\frac{H-\sum_{t\in\mathcal{I}}\mu_t}{\sqrt{\sum_{t\in\mathcal{I}}\mu_t(1-\mu_t)}}.
\end{equation}
$Z_{\mathrm{cal}}$ subtracts what the clean model would have produced, such that a predictable token that happens to be green no longer counts as evidence. The standard KGW test~\cite{kirchenbauer2023watermark} instead compares $H$ with $\gamma N_{\mathrm{trig}}$, so its null distribution shifts with how predictable each text is, which widens the spread of innocent scores and raises the threshold needed for an FPR.

The adversary controls the sampling from the stolen model, for instance at a lower temperature, but the defender has no knowledge of this beforehand. We therefore compute $\mu_t$ from the clean model's unmodified next-token distribution at temperature 1, which needs no knowledge of the decoding and does not tie the detector to one temperature. The green list is drawn independently of all text, so a decoder that does not know it cannot favor green tokens, whatever its temperature. Averaged over the draw of the list, innocent text therefore collects as many green hits as $\sum_{t\in\mathcal{I}}\mu_t$ predicts, whether it was sampled at temperature 0.3, 0.7, or 1, or decoded greedily, and only a model that carries the watermark exceeds this count systematically. For text sampled from the clean model at temperature 1, the case for which $Z_{\mathrm{cal}}$ is constructed, this holds at every position and for every fixed list. For one fixed list, the scores of text sampled at other temperatures can still spread differently around zero, as can those of human text and text of another model. The empirical threshold below absorbs this spread for the kinds of text it is calibrated on, and Section~\ref{subsec:deployment} tests the most extreme decoding, greedy, on watermarked and innocent text. The decision uses the score of the whole suspect text. To measure how detection grows with length, we also score its prefixes in steps of 50 tokens, each prefix length with its own threshold (Appendix~\ref{subsec:seq_len}). Since tokens and routing events within a text are not independent, we do not interpret $Z_{\mathrm{cal}}$ as a calibrated $p$-value. Instead, we calibrate the decision threshold empirically on innocent text.

\subsection{Decision Threshold}
\label{subsec:thresholds}

The next question is which innocent text the threshold should be calibrated on. A provenance claim has to hold for any innocent text the defender may be shown, not only for the text that is easiest to collect. The clean model's own generations, the natural null distribution for a model-specific watermark, do not suffice because they are the easiest innocent text for this detector. $Z_{\mathrm{cal}}$ is centered on that model, so the clean model's own text scores near zero, whereas human text, text of another model family, and text of a fine-tuned derivative of the same base model spread wider. A threshold set on the clean model's text alone therefore flags these sources more often than the target rate (Section~\ref{subsec:false_positives}).

\ourMethod therefore calibrates the threshold on a set of null sources $\{\mathcal{D}_s\}$, each scored by the same detector as the suspect text. We use four: (1) the clean model's own text, (2) human text, (3) text of another model family, and (4) text of a benign fine-tune of the same base model, the closest innocent relative of the watermarked model, which shares its base and tokenizer but not its watermark. For a target FPR $\alpha$, the threshold is the largest of the sources' empirical $(1-\alpha)$-quantiles of $Z_{\mathrm{cal}}$,
\begin{equation}\label{eq:threshold}
\tau=\max_{s}\,\hat{Q}_{1-\alpha}\big(\{Z_{\mathrm{cal}}(x):x\in\mathcal{D}_s\}\big).
\end{equation}
A text is flagged when its score exceeds $\tau$. Since $\tau$ is at least each source's own quantile, at most a fraction $\alpha$ of each source's calibration texts is flagged. A single quantile over the pooled sources would only bound their average, which depends on how many texts of each kind the pool contains. We use $\alpha=1\%$, the standard operating point of recent watermarking work~\cite{gloaguen2026unifiedframeworkllmwatermarks,gloaguen2026diffusionwatermarking}.


\textbf{Held-Out Calibration.} Bounding the FPR on the calibration texts is not enough to claim the same FPR for the texts a defender will audit later. A threshold evaluated on the texts it was set on yields an FPR of at most 1\% by construction, whatever its rate on new text would be, so this number carries no evidence about real deployment, where every audited text is one the threshold has never seen. We therefore set every threshold on a tune split of the null sources and measure the TPR and the FPR of every source on a disjoint test split (Section~\ref{subsec:datasets}), whose texts are exchangeable with the tune texts by construction; selecting and evaluating on the same data would bias the reported rates.

\section{Implementation and Evaluation Setup}
\label{sec:implementation}


\subsection{Target Models}
\label{subsec:target_models}

We evaluate \ourMethod on eight open-source MoE LLMs that span a diverse set of architecture types, number of experts and layers, parameter counts, developers, and top-$k$ activation values: DeepSeek-V2-Lite-Chat~\cite{deepseek2024deepseekv2}, GPT-OSS-20B~\cite{openai2025gptossmoe}, Hunyuan-A13B-Instruct~\cite{tencent2025hunyuanmoe}, Mixtral-8x22B-Instruct-v0.1~\cite{jiang2024mixtralmoe}, Nemotron-3-Nano-30B-A3B~\cite{nvidia2025nemotron3nanoopen}, OLMoE-1B-7B-0125-Instruct~\cite{muennighoff2024olmoe}, Phi-3.5-MoE-Instruct~\cite{abdin2024phi3moe}, Qwen3.6-35B-A3B~\cite{alibaba2024qwen36moe}. The relevant model architecture specifications are detailed in Table~\ref{tab:model_specifications}. Below, we use their family names (e.g., OLMoE and DeepSeek).

\begin{table*}[!t]
  \centering
  \footnotesize
  \caption{Architecture specifications of the targeted MoE LLMs with the number of candidate experts.}
  \label{tab:model_specifications}
  \begin{tabular}{lccccccc}
    \toprule
    & \textbf{Sparse} & \textbf{Shared} & \textbf{MoE} &  & \textbf{Active/Total} &  & \textbf{Candidate}\\
    \textbf{Model} & \textbf{Experts} & \textbf{Experts} & \textbf{Layers} & \textbf{Top-$k$} & \textbf{Params} & \textbf{Developer} & \textbf{Experts}\\
    \midrule
    DeepSeek-V2-Lite-Chat       & 64  & 2   & 26 & 6 & 2.4B / 15.7B  & DeepSeek  & 169 \\
    GPT-OSS-20B                 & 32  & N/A & 24 & 4 & 3.6B / 21B    & OpenAI    & 84 \\
    Hunyuan-A13B-Instruct       & 64  & 1   & 32 & 8 & 13B / 80.4B   & Tencent   & 208 \\
    Mixtral-8x22B-Instruct-v0.1 & 8   & N/A & 56 & 2 & 39B / 141B    & Mistral   & 28 \\
    Nemotron-3-Nano-30B-A3B     & 128 & 2   & 23 & 6 & 3.5B / 30B    & NVIDIA    & 300 \\
    OLMoE-1B-7B-0125-Instruct   & 64  & N/A & 16 & 8 & 1.3B / 7.1B   & AI2       & 104 \\
    Phi-3.5-MoE-Instruct        & 16  & N/A & 32 & 2 & 6.6B / 41.9B  & Microsoft & 48 \\
    Qwen3.6-35B-A3B             & 256 & 1   & 40 & 8 & 3B / 35B      & Alibaba   & 1,020 \\
    \bottomrule
  \end{tabular}
\end{table*}

\subsection{Datasets}
\label{subsec:datasets}

To profile expert activation frequencies for target selection (Section~\ref{subsec:payload_creation}), we process 4,000 randomly sampled paragraphs from the WikiText-2~\cite{merity2016pointer} training split, each truncated to 512 tokens, and count expert activations across MoE layers. This profiling dataset is kept separate from all evaluation data. For watermark injection, we construct a pool of 8,000 training texts from the C4~\cite{raffel2020c4} and ELI5~\cite{fan2019eli5} training splits, alternating between C4 documents and ELI5 questions. All evaluation prompts and reference texts are excluded from training to prevent data leakage.

We evaluate \ourMethod on C4 and ELI5, following established watermarking benchmarks~\cite{kirchenbauer2023watermark,he2025theoretically,gloaguen2026unifiedframeworkllmwatermarks}. C4 represents open-ended text continuation, while ELI5 evaluates long-form question answering. For C4, we use the first 50 tokens as prompts and evaluate the generated continuations; for ELI5, we use questions of up to 150 tokens and evaluate the generated answers. All prompts are provided without chat templates. Each dataset is shuffled and divided into disjoint tune and test splits. The tune split uses 100 prompts per dataset (200 texts) to select the watermark strength $\delta$, while the test split uses 500 prompts per dataset (1,000 texts) for final evaluation. Each null source provides approximately 1,000 texts per split (Section~\ref{subsec:null_sources}). Its tune texts set the detection thresholds and its test texts give the held-out FPR, so neither $\delta$ nor any threshold is chosen on the texts we report.





\subsection{Payload \& Watermark Injection}
\label{subsec:payload_details}

Each payload is derived from a secret seed $K$, which determines the green list, target layers, and target experts. We evaluate three seeds per model, with each seed producing model-specific target configurations. The green-list fraction is fixed at $\gamma=0.25$, following prior watermarking work~\cite{kirchenbauer2023watermark,gu2024learnability}.
For a model with $L$ MoE layers indexed from 0, we select three target layers from the second half ($\lfloor L/2\rfloor$ to $L-1$). These candidate layers are divided into three contiguous, approximately equal-sized bins, and one layer is sampled uniformly from each bin using $K$.

Within each selected layer $l$, we sample one target expert from the 40th--60th percentile of activation frequency, estimated using the profiling dataset (Section~\ref{subsec:datasets}). Let $c_{l,e}$ denote the activation count of expert $e$ in layer $l$, and $r_l(e)\in\{0,\dots,N-1\}$ its ascending rank among the $N$ routed experts. The candidate expert band is defined as
\begin{equation}\label{eq:band}
\mathcal{B}_l=\big\{e:\lfloor 0.4N\rfloor\le r_l(e)\le\lfloor 0.6N\rfloor-1\big\},
\end{equation}
from which the target expert $e_l^*$ is sampled uniformly using $K$.
Table~\ref{tab:model_specifications} reports the total number of candidate experts for each model. We conservatively assume that the adversary knows this candidate set, although identifying it precisely may require access to the clean model's routing profile. As shown in Section~\ref{subsec:broad_masking}, masking all candidates incurs substantial model utility costs.

We train rank-16 LoRA adapters with $\alpha_{\text{LoRA}}=$ 32 and dropout 0.05 on the gate, up, and down projections of the three target experts for 600 steps, at a constant learning rate of 10$^{-\text{3}}$ without warm-up, with batches of 8 sequences of at most 512 tokens and training seed 0. We use the same settings for every model, seed, and $\delta$ and do not tune them per model. Across the eight models, the adapters train at most 0.006\% of a model's parameters, 0.37M to 3.24M depending on the model, against 156M to 1.28B for the benign fine-tune (Section~\ref{subsec:null_sources}).


\subsection{Null Sources}
\label{subsec:null_sources}

For each model and split, each of the four null sources (Section~\ref{subsec:thresholds}) holds about 1,000 texts, which we score with the detector of the watermarked model under evaluation. The clean model's texts are its generations for the evaluation prompts of the split. The human texts are the C4 continuations and ELI5 answers of the same prompts, truncated to 300 tokens. The other family's texts are the clean texts of another evaluated model, retokenized with the audited model's tokenizer and cut at 300 tokens; we generate them with 400 tokens so that they still reach 300 after retokenization. The benign fine-tune's texts come from the clean model fine-tuned with a recipe that does not depend on the seed or the target experts. The benign fine-tuning procedure trains rank-16 LoRA adapters on every linear layer except the output layer with the plain language-modeling loss, on the same training texts and for the same 600 steps as watermark injection, at a learning rate of 2$\times\text{10}^{-\text{4}}$.

\subsection{Evaluation Metrics}
\label{subsec:evaluation_metrics}


\textbf{Watermark Detectability.} We frame watermark detection as a constrained hypothesis test and report the TPR at 1\% FPR, a standard adopted in recent watermarking literature~\cite{gloaguen2026unifiedframeworkllmwatermarks,gloaguen2026diffusionwatermarking}, with our worst-case threshold over the four null sources (Section~\ref{subsec:thresholds}). Next to the TPR, we report the held-out FPR of each source. Each source's threshold is an empirical quantile of about 1,000 tune texts, so its rate on held-out texts carries sampling error and can fall on either side of the 1\% target; we report these rates as measured.

\textbf{Generation Quality.} We measure generation quality with oracle perplexity. Perplexity measures how predictable a text is to a fixed oracle model, so a watermark that distorts the model's next-token distribution raises the perplexity of its text. Following related work~\cite{gloaguen2026unifiedframeworkllmwatermarks}, a fixed oracle model, Qwen3-30B-A3B-Instruct-2507~\cite{qwen3technicalreport}, scores each generated text given its prompt, and reports the relative increase in perplexity over the clean model on the same prompts, $\Delta$PPL. 

\textbf{Benchmark Accuracy.} While perplexity measures shifts of the output distribution, it evaluates predictability rather than whether the model still solves tasks. To evaluate whether watermarking preserves the model's capabilities, we measure accuracy on GSM8K~\cite{cobbe2021gsm8k}, HumanEval~\cite{chen2021codex}, IFEval~\cite{zhou2023instruction}, MMLU~\cite{hendrycks2021mmlu}, and WinoGrande~\cite{sakaguchi2020winogrande}, which cover mathematical reasoning, code generation, instruction following, broad world knowledge, and commonsense reasoning, respectively.


\section{Experimental Results}
\label{sec:experimental_results}

\subsection{Optimizing Watermark Strength}
\label{subsec:watermark_strength}

The strength $\delta$ trades detection against utility. We fix one operating point $\delta^*$ per model with a rule that sees only the tune split, and examine how detection and both utility measures behave around it.

\textbf{Selection Rule.} For every model, we inject the watermark with $\delta\in\{1,2,3,4,6,8\}$ and take the knee of its curve of TPR against $\Delta$PPL with Kneedle~\cite{satopaa2011kneedle}, the point farthest above the diagonal once both axes are scaled to $[0,1]$. Because our rule reads only the tune split, we never tune $\delta$ on the test texts, so the held-out results of Table~\ref{tab:main_results} are not biased by the choice of the operating point. This reflects real deployment, where a defender fixes the strength before seeing any suspect text.

\begin{figure}[!t]
  \centering
  \includegraphics{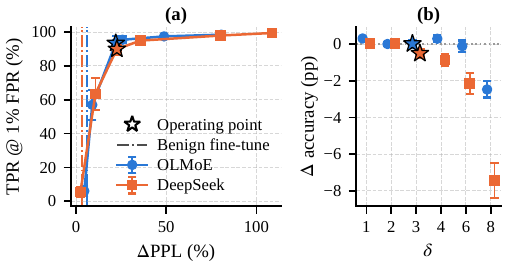}
    \caption{Impact of watermark strength $\delta$ on detection and model utility. (a) TPR at 1\% FPR (\%) versus perplexity increase (\%), with stars marking the selected operating points $\delta^*$ and dash-dotted lines marking the benign fine-tune. (b) Change in mean benchmark accuracy (pp). Results are averaged over three seeds, with error bars showing standard deviation.}
  \label{fig:delta_sweep}
\end{figure}

\textbf{Detectability--Utility Trade-off.} Figure~\ref{fig:delta_sweep} shows detection against perplexity and benchmark accuracy on the tune split for OLMoE and DeepSeek, and Table~\ref{tab:main_results} lists $\delta^*$ for every model. Detection rises steeply up to $\delta=3$ and then saturates, while the perplexity cost keeps growing (Figure~\ref{fig:delta_sweep}a), so the rule selects $\delta^*=3$ for both models; a moderate strength captures most of the attainable detection, and stronger watermarks buy little more at a steep cost. Benchmark accuracy reacts only at higher values of $\delta$ than perplexity (Figure~\ref{fig:delta_sweep}b). Up to $\delta = 4$, it changes by at most 0.9\% on both models, and larger losses appear only at $\delta = 6$ and 8 on DeepSeek (2.1 and 7.4\%) and at $\delta = 8$ on OLMoE (2.5\%). The operating point thus lies below the strengths at which accuracy drops, and perplexity, the more sensitive of the two measures, is the right cost axis for our selection rule. However, perplexity alone is not enough; the benign fine-tune of each model (dash-dotted lines) costs less perplexity than the watermark at any $\delta$ above 1, yet loses more benchmark accuracy than the watermark at $\delta^*$ (Table~\ref{tab:benchmarks}), so we report both measures throughout.

\subsection{Detection and Utility at the Operating Point}
\label{subsec:main_results}

\textbf{Detection.} With $\delta^*$ fixed on the tune split, we measure detection and its cost on the test split, on which neither the operating point nor any threshold was chosen. Table~\ref{tab:main_results} reports the mean and standard deviation over the three seeds of each model. At 1\% FPR under the worst-case threshold, \ourMethod detects, on average, 92.5\% of watermarked texts from 300 tokens of text alone, from 88.4\% on Qwen to 95.3\% on OLMoE. The three seeds, which place the watermark in different experts and draw different green lists, give standard deviations of 2.9 to 4.6\%.

\textbf{False Positives.} Detection is only useful if innocent text is rarely flagged. Table~\ref{tab:main_results} lists the held-out FPR of human text, the other family's text, and the benign fine-tune's text, while the clean model's own text, the easiest source, is flagged at 0.0\% on OLMoE and 0.3\% on DeepSeek on average. Each threshold is an empirical 99th percentile of about 1,000 tune texts per source, so the rate on held-out texts scatters around the target. Averaged over seeds, the held-out FPR of every source in Table~\ref{tab:main_results} lies between 0.4 and 1.3\%. Section~\ref{subsec:false_positives} shows why the threshold has to cover every source.

\textbf{Generation Quality.} The watermark raises the oracle perplexity of the generated text by 11.6 to 16.0\% (Table~\ref{tab:main_results}). The benign fine-tune, trained on the same texts for the same number of steps, costs less perplexity than the watermark (Figure~\ref{fig:delta_sweep}a) but, as the next paragraph shows, more benchmark accuracy.

\textbf{Benchmark Accuracy.} The mean accuracy over the five benchmarks changes by between $-$0.2 and $-$1.1\% (Table 3), $-$0.4\% on average, so the operating point costs little accuracy; only Mixtral loses more than 0.6\%. Table~\ref{tab:benchmarks} in Appendix~\ref{app:per_benchmark_accuracy} reports the accuracy per benchmark. Across all models, \ourMethod stays within 0.8\% of the clean model on MMLU, GSM8K, WinoGrande, and IFEval, while losing 1.0 to 3.5\% on HumanEval ($-1.8$\% on average). This difference is consistent with code generation requiring precise syntactic token choices that are more easily disrupted by a vocabulary bias~\cite{lee2024codewatermarking}. Conversely, the benign fine-tune preserves HumanEval ($-$0.6\% on average) but loses 2.5 to 4.3\% on GSM8K ($-$3.4\% on average) and 1.1 to 2.6\% on IFEval ($-$1.9\% on average).

\begin{table*}[!t]
  \centering
  \footnotesize
  \caption{Detection and utility at the operating point $\delta^*$ on the test split, reported as mean $\pm$ SD over three seeds. TPR is measured at 1\% FPR using the worst-case threshold across four null sources; FPRs are reported separately for three null sources. Utility is measured by perplexity increase (\%) and mean benchmark accuracy change (pp) relative to the clean model.}
  \label{tab:main_results}
  \begin{tabular}{lccccccc}
    \toprule
    &  &  & \multicolumn{3}{c}{FPR (\%)} &  &  \\
    \cmidrule(lr){4-6}
    Model & $\delta$ & TPR (\%) & Human & Other family & Benign & $\Delta$PPL (\%) & $\Delta$Acc. (pp) \\
    \midrule
    DeepSeek-V2-Lite-Chat       & 3 & 91.6 $\pm$ 3.7 & 1.2 $\pm$ 0.9 & 0.4 $\pm$ 0.4 & 1.2 $\pm$ 0.5 & +12.8 $\pm$ 6.0 & $-$0.5 $\pm$ 0.2 \\
    GPT-OSS-20B                 & 3 & 94.1 $\pm$ 3.2 & 0.9 $\pm$ 0.4 & 0.6 $\pm$ 0.5 & 0.8 $\pm$ 0.4 & +15.1 $\pm$ 2.8 & $-$0.3 $\pm$ 0.2 \\
    Hunyuan-A13B-Instruct       & 3 & 92.8 $\pm$ 3.5 & 1.1 $\pm$ 0.6 & 0.5 $\pm$ 0.3 & 1.0 $\pm$ 0.5 & +13.4 $\pm$ 4.1 & $-$0.3 $\pm$ 0.2 \\
    Mixtral-8x22B-Instruct-v0.1 & 2 & 92.2 $\pm$ 2.9 & 0.7 $\pm$ 0.3 & 0.5 $\pm$ 0.4 & 0.9 $\pm$ 0.3 & +14.2 $\pm$ 2.4 & $-$1.1 $\pm$ 0.6 \\
    Nemotron-3-Nano-30B-A3B     & 4 & 91.9 $\pm$ 4.1 & 1.0 $\pm$ 0.5 & 0.7 $\pm$ 0.5 & 1.1 $\pm$ 0.6 & +13.9 $\pm$ 4.8 & $-$0.4 $\pm$ 0.3 \\
    OLMoE-1B-7B-0125-Instruct   & 3 & 95.3 $\pm$ 4.3 & 0.8 $\pm$ 0.2 & 0.8 $\pm$ 0.6 & 0.5 $\pm$ 0.3 & +16.0 $\pm$ 1.3 & $-$0.2 $\pm$ 0.2 \\
    Phi-3.5-MoE-Instruct        & 3 & 93.4 $\pm$ 3.6 & 0.9 $\pm$ 0.5 & 0.6 $\pm$ 0.4 & 0.7 $\pm$ 0.4 & +14.7 $\pm$ 3.1 & $-$0.4 $\pm$ 0.3 \\
    Qwen3.6-35B-A3B             & 4 & 88.4 $\pm$ 4.6 & 1.3 $\pm$ 0.8 & 0.6 $\pm$ 0.5 & 1.1 $\pm$ 0.5 & +11.6 $\pm$ 5.2 & $-$0.3 $\pm$ 0.2 \\
    \midrule
    \textit{Average} &   & \textit{92.5} & \textit{1.0} & \textit{0.6} & \textit{0.9} & \textit{+14.0} & \textit{$-$0.4} \\
    \bottomrule
  \end{tabular}
\end{table*}

\subsection{False Positives Across Null Sources}
\label{subsec:false_positives}

\textbf{Clean-Only Threshold.} Calibrating on four sources raises the threshold and costs detection, so we first check whether a threshold calibrated on the clean model's text alone would already keep the FPR of every source near 1\%. $Z_{\mathrm{cal}}$ centers innocent text of every kind near zero, but human text, the other family's text, and the benign fine-tune's text spread wider than the clean model's own text, with standard deviations of 1.0 to 1.3 against about 0.9. A threshold set on the clean model's text alone therefore falls inside their upper tails, as Figure~\ref{fig:zcal_scores_appendix} in Appendix~\ref{app:statistics} shows. Averaged over seeds, this clean-only threshold flags 4.7\% of human text, 4.1\% of the other family's text, and 4.5\% of the benign fine-tune's text on OLMoE, and 2.9, 1.8, and 3.2\% on DeepSeek, while it flags about 0.9\% of the clean model's own text on both. It thus misses the 1\% target on every source except the one it was calibrated on.

\textbf{Worst-Case Threshold.} For each null source, we take the $Z_{\mathrm{cal}}$ score that 99\% of its tune texts stay below, and our threshold is the largest of these four scores (Equation~\ref{eq:threshold}). Which source gives the largest score differs between watermarked models. Among the six seeds of OLMoE and DeepSeek, it is the benign fine-tune for three seeds, human text for two, and the other family for one. A threshold calibrated on any single source would exceed the 1\% target on the other sources for some seeds, whereas calibrating on all four keeps every source near the target (Section~\ref{subsec:main_results}), at a modest cost in TPR against the clean-only threshold.

\subsection{Comparison with Weight-Embedded Baselines}
\label{subsec:experiments_comparison}

\ourMethod is a watermark that survives the theft of the weights, so we compare its cost with watermarks that operate in the same threat model.

\textbf{Baselines.} KGW-D~\cite{gu2024learnability} distills the KGW distribution into all of a model's weights. We train it using LoRA adapters on every linear layer and the KGW target at every position. GaussMark~\cite{block2025gaussmark} adds Gaussian noise to selected weights and detects it through the gradient of the suspect text's log-likelihood with respect to the perturbed matrices, computed on the defender's clean weights.

\textbf{GaussMark Variants.} GaussMark was designed for dense models, in which one feed-forward matrix acts on every token. An MoE layer has no such matrix, because its feed-forward network is split into experts that each see only some of the tokens, so GaussMark has no single faithful port to MoE models. We therefore evaluate two variants. GaussMark$_{\mathrm{D}}$ keeps the dense positioning and perturbs the up projection of every expert in one late layer, shared experts included, so that every token still passes through perturbed weights. The targeted layer follows GaussMark's original setting criteria. GaussMark$_{\mathrm{E}}$ adopts the sparsity of \ourMethod. It perturbs exactly the parameters that \ourMethod changes and differs only in how the watermark is embedded, as random noise instead of a trained bias, and how it is detected, through the log-likelihood gradient instead of replayed routing.


\textbf{Results.} Table~\ref{tab:baselines} compares the four watermarks. All methods reach a similar FPR. KGW-D and GaussMark$_{\mathrm{D}}$, which act on every token, reach a TPR of 85.8 to 90.1\%, below that of \ourMethod (95.3 and 91.6\%), raise perplexity 21.4 to 32.4\% more than \ourMethodNoSpace, and cost 2.9 to 5.8\% benchmark accuracy, whereas \ourMethod reduces accuracy by at most 0.5\%. \ourMethod is thus better than both baselines in detection, generation quality, and benchmark accuracy on both models, so it Pareto-dominates them. GaussMark$_{\mathrm{E}}$, confined to the target experts, costs little perplexity (4.6 and 3.9\%) and almost no accuracy (+0.1 and $-$0.1), but detects only 46.3 and 38.7\% of watermarked texts. The advantage of \ourMethod also extends beyond utility to robustness, since under adversarial fine-tuning it retains more of its detection than both baselines (Section~\ref{subsec:fine_tuning}).

\begin{table}[!t]
  \centering
  \footnotesize
  \setlength{\tabcolsep}{4.9pt}
  \caption{\ourMethod against weight-embedded watermarks on the test split. Columns give the TPR at 1\% FPR (\%), the highest FPR (\%) of a null source, the perplexity increase (\%) over the clean model, and the change in mean benchmark accuracy (pp). Values are means over the three seeds of each model.}
  \label{tab:baselines}
  \begin{tabular}{lcccc}
    \toprule
    Method & TPR (\%) & Max FPR (\%) & $\Delta$PPL (\%) & $\Delta$Acc. (pp) \\
    \midrule
    \multicolumn{5}{l}{\textit{DeepSeek}} \\
    WoE (ours) & 91.6 & 1.2 & $+$12.8 & $-$0.5 \\
    KGW-D & 85.8 & 1.3 & $+$34.2 & $-$4.8 \\
    GaussMark$_{\mathrm{D}}$ & 86.3 & 1.1 & $+$43.8 & $-$2.9 \\
    GaussMark$_{\mathrm{E}}$ & 38.7 & 1.0 & $+$3.9 & $-$0.1 \\
    \midrule
    \multicolumn{5}{l}{\textit{OLMoE}} \\
    WoE (ours) & 95.3 & 0.8 & $+$16.0 & $-$0.2 \\
    KGW-D & 90.1 & 1.1 & $+$48.4 & $-$5.8 \\
    GaussMark$_{\mathrm{D}}$ & 88.1 & 0.9 & $+$37.6 & $-$3.3 \\
    GaussMark$_{\mathrm{E}}$ & 46.3 & 0.8 & $+$4.6 & $+$0.1 \\
    \bottomrule
  \end{tabular}
\end{table}

\subsection{Layer Selection}
\label{subsec:ablation_layer}

\textbf{Ablation Protocol.} The default layer selection rests on the reasoning that late layers form the next-token distribution and that a change made in an early layer would have to survive every later layer (Section~\ref{subsec:payload_creation}). To test this reasoning, and to measure how many target layers the watermark needs, we vary one setting at a time from the default and keep everything else fixed: $\delta^*$, the training steps, the LoRA rank, and, for each of the three seeds, the green list of its default configuration. Each configuration is evaluated on the tune split. 
We report the TPR and the mean $Z_{\mathrm{cal}}$ where the TPR saturates. Table~\ref{tab:ablation} in Appendix~\ref{app:ablations} shows every level.

\textbf{Depth.} With the candidate range in the first half of the MoE layers, the TPR is 2.8\% on OLMoE and 4.8\% on DeepSeek, against 93.5 and 90.0\% for the default, and with the range in the last quarter, it is 94.2\% and 91.3\% at a perplexity cost of 19.8 and 20.8\%. The last quarter also reduces the number of candidate experts an adversary has to mask from 104 to 52 on OLMoE, and fixes layers 14 and 15 for every seed, so it leaves the seed less to hide. Targets in the first half are selected about as often as later ones, with trigger rates of 0.39 and 0.32 against 0.37 and 0.35, yet barely learn the bias. The watermark therefore needs late layers, while the last quarter matches the default and differs only in what the seed hides.

\textbf{Number of Target Layers.} One, three, or six target layers trade the trigger rate against the share of the model that carries the watermark. With one layer, the trigger rate falls to 0.12 on OLMoE and 0.11 on DeepSeek, and the single target barely learns the bias (mean $Z_{\mathrm{cal}}$ of 0.38 and 0.51), so the TPR is 0.0\% and 2.5\%. Six layers raise the TPR only to 94.8\% and 92.3\%, at a perplexity cost of 27.1\% and 29.4\% and an accuracy change of $-$0.5 and $-$0.9\%. The candidate experts stay the same at every level, so the number of layers changes only the detection and cost, and three layers capture nearly all of the detection at a lower cost than six.

\subsection{Expert Selection}
\label{subsec:ablation_expert}

\textbf{Band Position.} Target experts from the 0--20\% band trigger at a rate of 0.28 on OLMoE and 0.22 on DeepSeek and reach a TPR of only 13.0 and 16.8\%, against 93.5 and 90.0\% for the default. Target experts from the 80--100\% band trigger about as often as the default (0.35 and 0.33) at a lower perplexity cost (14.1 and 13.8\%), but reach only 51.0 and 44.7\%. Both ends thus carry a far weaker watermark, consistent with the reasoning that rare experts see too few positions to learn the bias and frequent ones serve inputs too varied for a low-rank update to bias them (Section~\ref{subsec:payload_creation}).

\textbf{Experts per Layer.} Two or eight target experts per layer raise the trigger rate to 0.58 and 0.95 on OLMoE and the perplexity cost to 28.1 and 49.3\%, but the TPR only to 93.8 and 97.8\%, against 93.5\% for one expert, and DeepSeek behaves similarly. With eight experts, nearly every token passes through a target expert, so the watermark is no longer sparse, and the model loses 1.0 and 1.5\% accuracy. One expert per layer therefore keeps the watermark sparse at almost no loss in detection.

\textbf{Band Width.} The width of the band changes only what the seed hides, so its effect follows from the routing profile without training. Widening the band from 45--55\% through the default to the whole ranking raises the number of candidate experts an adversary without the seed has to mask from 56 through 104 to 512 on OLMoE and from 91 through 169 to 832 on DeepSeek. A wider band therefore makes blind masking costlier (Section~\ref{subsec:broad_masking}), at the price of targets that trigger more or less often than the default, as the band positions show. 
A wider band also increases the number of candidates for weight-based localization, although none of the three evaluated signals reliably identifies the targets even within the default band (Section~\ref{subsec:localization}).

\section{Security Evaluation}
\label{sec:security_evaluation}

The results so far assume an adversary who runs the stolen weights as they are. The threat model (Section~\ref{sec:threat_model}) assumes an adversary who knows the algorithm, holds the full watermarked checkpoint, and can change the weights or the text before publishing, without knowing the seed. We evaluate each attack by the TPR that remains at 1\% FPR on the test split under the thresholds of the main results and, where we measure it, by what the attack costs the adversary in perplexity and benchmark accuracy. An attack that removes the watermark only by ruining the model has not removed it in practice. Attacks that also change innocent text, such as quantization, different decoding, paraphrasing, or distillation, get a false-positive control with innocent text under the same transform, audited at the same thresholds.

\subsection{Masking the Candidate Experts}
\label{subsec:broad_masking}

\begin{figure}[!t]
  \centering
  \includegraphics{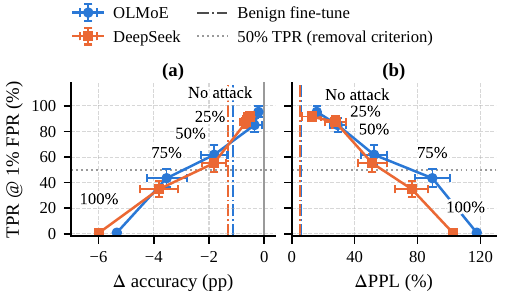}
  \caption{Impact of candidate-expert masking on watermark detection and model utility. TPR at 1\% FPR (\%) versus (a) change in mean benchmark accuracy (pp) and (b) perplexity increase (\%) over the clean model. Results are averaged over three seeds, with error bars showing standard deviation. Dash-dotted lines mark the cost of the benign fine-tune, while the dotted line marks the 50\% TPR removal threshold.}
  \label{fig:a2_masking}
\end{figure}

An adversary who knows the algorithm knows the candidate experts (Section~\ref{subsec:payload_details}) but not which of them carry the watermark, and it cannot test a guess without the green list. Its blind attack is therefore to reroute the candidates, so that the routers no longer select them and their tokens go to other experts. We mask random shares of 25, 50, and 75\% of the candidate experts and then all of them (104 of the 512 experts in OLMoE's candidate layers and 169 of the 832 in DeepSeek's), and measure the detectability and utility of each masked model (Figure~\ref{fig:a2_masking}). We measure that cost mainly by benchmark accuracy, because a stolen model is only worth keeping if it still solves tasks, and perplexity alone can mistake looping text for good text. Before running the attack, we fixed when it counts as removing the watermark, namely when the TPR falls below 50\% at an accuracy cost no larger than that of the benign fine-tune, a cost an adversary would plausibly accept. Because the masked model does not depend on the seed when every candidate is masked, its cost is that of the clean model with the same experts masked.

Masking removes the watermark only at a cost in model utility (Figure~\ref{fig:a2_masking}). Rerouting a random 25\% of the candidates leaves a TPR of 85.0\% on OLMoE and 87.4\% on DeepSeek, and rerouting 50\% leaves 61.8 and 55.2\% while already costing 1.8\% accuracy on both models, more than the benign fine-tune's 1.1 and 1.3\%. The TPR falls below 50\% only when 75\% of the candidates are masked (43.4 and 34.8\%), at a cost of 3.5 and 3.8\% accuracy and a perplexity increase of 89.4 and 76.4\%. Masking every candidate removes the watermark (0.8 and 0.7\%) but costs 5.3 and 6.0\% accuracy and raises the perplexity of the model's text by 117.7 and 102.5\%. No masked share therefore meets our removal criterion, since each either leaves the TPR above 50\% or costs more accuracy than the benign fine-tune.

\subsection{Localizing the Targets from the Weights}
\label{subsec:localization}

\begin{figure}[!t]
  \centering
  \includegraphics{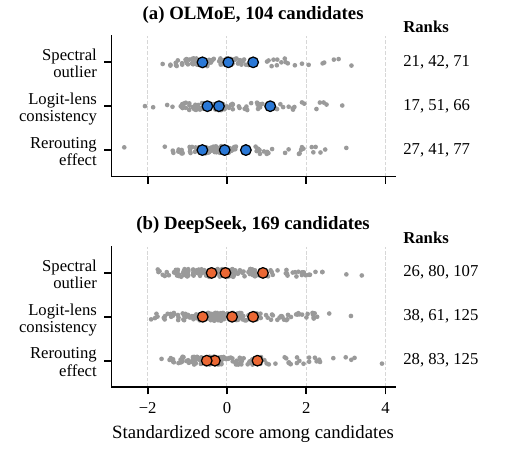}
  \caption{Localizing the target experts from the weights alone, for one seed. Standardized score of each candidate expert under the three signals: (a) the 104 of OLMoE and (b) the 169 of DeepSeek, with the three true targets marked and their ranks among the candidates on the right.}
  \label{fig:a3_localization}
\end{figure}

A stronger adversary first looks for the targets. Injection changes only three experts, so their weights or their effect on the output may stand out among the candidates. We score every candidate expert with three signals that need nothing but the watermarked checkpoint. The first is how much rerouting the expert changes the next-token distribution, the second how consistently the logit lens~\cite{nostalgebraist2020logitlens} reads a push toward the same tokens from its output, and the third how far its singular-value spectrum stands out from those of the other experts in its layer, which a rank-16 update may shift. The adversary then masks the candidates in order of suspicion (Figure~\ref{fig:a3_localization}).

None of the three signals singles out the targets (Figure~\ref{fig:a3_localization}). On OLMoE, the highest-ranked target lies at rank 17 of 104 (logit-lens consistency), and on DeepSeek at rank 26 of 169 (spectral outlier), and under every signal the three targets spread over the middle of the candidates. An adversary who masks candidates in order of suspicion therefore reaches the first target only after 17 and 26 experts, and all three only after 66 and 107, more than 60\% of the candidates, which costs model utility (Section~\ref{subsec:broad_masking}).

\subsection{Watermark Removal by Fine-Tuning}
\label{subsec:fine_tuning}

\begin{figure}[!t]
  \centering
  \includegraphics{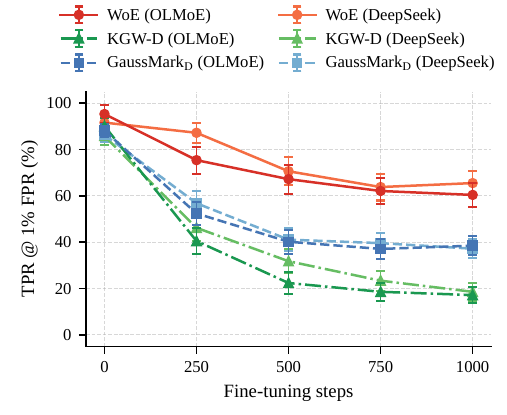}
  \caption{Removal by fine-tuning, as the TPR at 1\% FPR (\%) on the test split against fine-tuning steps for \ourMethod and the weight-embedded baselines. Points are means over the three seeds of each model, and bars show one standard deviation.}
  \label{fig:fine_tuning}
\end{figure}

An adversary can fine-tune the stolen model on ordinary text and hope to overwrite the target experts without knowing them. We continue training each watermarked model with the recipe of the benign fine-tune (Section~\ref{subsec:null_sources}) on OpenWebText~\cite{gokaslan2019openweb}, a public corpus of the adversary's choosing. We fine-tune at the benign learning rate of $2\times10^{-4}$ and audit checkpoints at 250, 500, 750, and 1,000 steps on the test split with the thresholds of the main results. KGW-D and GaussMark$_{\mathrm{D}}$ undergo the same fine-tuning procedure (Figure~\ref{fig:fine_tuning}). KGW-D is known to fade under further fine-tuning on ordinary text~\cite{gu2024learnability}.

\ourMethod fades more slowly than both baselines (Figure~\ref{fig:fine_tuning}). After 1,000 steps, it still detects 60.4\% of OLMoE's and 65.5\% of DeepSeek's watermarked texts, whereas KGW-D falls to 17.2 and 18.6\% and GaussMark$_{\mathrm{D}}$ to 38.5 and 37.2\%. Most of the loss occurs within the first 500 steps, after which \ourMethod levels off. Texts of the clean model fine-tuned with the same recipe are flagged at 0.8 to 1.2\% at every checkpoint, so fine-tuning does not raise false positives.

\subsection{Quantization and Decoding}
\label{subsec:deployment}

An adversary also decides how to deploy the stolen model. Quantizing it to int8 or int4 saves memory but changes the computation that the defender replays. The decoding temperature cannot shift the scores of innocent text systematically, since no decoding that ignores the green list favors green tokens (Section~\ref{subsec:watermark_detection}), but it changes how widely those scores spread and how strongly the watermark shows in the sampled tokens. Greedy decoding is the extreme of both, so we test it on watermarked text and, as its false-positive control, on innocent text decoded the same way. Table~\ref{tab:robustness} reports the residual TPR under each modification and the attacker's cost.

Quantization leaves the watermark in place (Table~\ref{tab:robustness}). Int8 lowers the TPR by at most 2.7\%, to 92.6\% on OLMoE and 90.9\% on DeepSeek, and int4 to 89.5 and 86.4\%, while int4 itself costs the adversary 1.1 and 1.4\% accuracy. Greedy decoding keeps most of the watermark (88.4 and 83.5\%) but breaks the false-positive control, since greedily decoded innocent text is flagged at up to 5.5 and 4.8\%. As Section~\ref{subsec:watermark_detection} anticipates, greedy decoding does not favor green tokens, but for one fixed list it widens the spread of innocent scores beyond what thresholds calibrated on sampled text absorb. A defender who expects greedy text therefore needs thresholds calibrated on greedy null text.

\begin{table}[!t]
  \centering
  \footnotesize
  \setlength{\tabcolsep}{5.5pt}
  \caption{Robustness to post-theft attacks at $\delta^*$ on the test split. TPR, perplexity change, and benchmark accuracy change are reported as the mean over three seeds; FPR is the maximum across seeds and transformed innocent-text sources. Utility changes ($\Delta$) are relative to the watermarked model. Dashes indicate metrics not applicable to text-only transformations.}
  \label{tab:robustness}
  \begin{tabular}{lcccc}
    \toprule
    Attack & TPR (\%) & FPR (\%) & $\Delta$PPL (\%) & $\Delta$Acc. (pp) \\
    \midrule
    \multicolumn{5}{l}{\textit{DeepSeek}} \\
    Int8 quant.       & 90.9 & 1.2 & $+$0.9 & $-$0.1 \\
    Int4 quant.     & 86.4 & 0.4 & $+$5.3 & $-$1.4 \\
    Greedy dec.       & 83.5 & 4.8 & \textemdash & \textemdash \\
    Paraphrasing            & 57.0 & 1.1 & \textemdash & \textemdash \\
    Distilled stud. & 81.1 & 1.3 & $+$21.0 & $-$3.1 \\
    \midrule
    \multicolumn{5}{l}{\textit{OLMoE}} \\
    Int8 quant.       & 92.6 & 0.8 & $+$0.8 & $-$0.1 \\
    Int4 quant.       & 89.5 & 0.2 & $+$4.1 & $-$1.1 \\
    Greedy dec.         & 88.4 & 5.5 & \textemdash & \textemdash \\
    Paraphrasing            & 34.9 & 1.0 & \textemdash & \textemdash \\
    Distilled stud. & 79.2 & 1.4 & $+$18.5 & $-$2.4 \\
    \bottomrule
  \end{tabular}
\end{table}

\subsection{Model Extraction by Distillation}
\label{subsec:distillation}

Instead of publishing text produced by the stolen weights, an adversary can distill them into a dense student and publish the student's text (Section~\ref{sec:threat_model}). A robust provenance mechanism should survive this extraction pipeline~\cite{sander2024watermarking,gu2024learnability,yi2025watermarkattack}, allowing the defender to use the text generated by a suspect, and possibly dense, student model to show that the student was trained on text generated by the defender's stolen weights.

We prompt the watermarked teacher with 40,000 prefixes of 50 tokens from OpenWebText~\cite{gokaslan2019openweb}, generate 512 tokens for each, and fine-tune the dense Llama-3-8B-Instruct~\cite{grattafiori2024llama3herdmodels} on this corpus with a standard LoRA recipe (rank 16, learning rate 2$\times\text{10}^{-\text{4}}$, cosine schedule, three epochs) that is not tuned for the watermark. The defender audits the student's text with its watermarked checkpoint and thresholds, and a student distilled from the clean model on the same prompts is the control. The student has no experts, so the watermark survives only if the student has learned to prefer green tokens in the contexts that trigger the target experts.

The watermark transfers to the student. Its texts are detected at 79.2\% with OLMoE as the teacher and 81.1\% with DeepSeek, whereas the texts of the control students are flagged at 1.4 and 1.3\% (Table~\ref{tab:robustness}). These control rates lie slightly above the 1\% target, so a defender who audits distilled students should include control students among its null sources. Distillation costs the adversary 18.5 and 21.0\% perplexity and 2.4 and 3.1\% accuracy relative to the watermarked teacher.

\subsection{Paraphrasing}
\label{subsec:text_perturbations}

Before publishing, an adversary can also rewrite the text, a standard attack on text watermarks~\cite{kirchenbauer2023watermark,kuditipudi2024robust,rastogi2024paraphrase}. A model from another family, Qwen2.5-7B-Instruct~\cite{qwen2025technicalreport}, paraphrases each watermarked test text, and we audit the paraphrase with the thresholds of the main results (Table~\ref{tab:robustness}). A paraphrase changes the tokens and, with them, the routing that the defender replays, so it attacks the triggered positions and the green-list hits at once. Innocent text paraphrased the same way is the false-positive control reported for this scenario. Additionally, we calculate the BERTScore F1~\cite{zhang2020bertscore} between each text and its paraphrase, rescaled with its authors' baseline so that unrelated text scores 0 and identical text 1. This checks that the paraphrase keeps the meaning.

Paraphrasing removes much of the watermark. The TPR falls to 34.9\% on OLMoE and 57.0\% on DeepSeek, while paraphrased innocent text is flagged at most 1.1\% of the time (Table~\ref{tab:robustness}). The paraphrases preserve the meaning of the original text, with a BERTScore F1 of 0.63 against the watermarked text, compared with 0.13 between two texts the model writes for the same prompt, so a paraphrase stays much closer to its original than another answer on the same topic does.
\section{Discussion}
\label{sec:discussion}

\textbf{Practical Deployability and Detection Overhead.} Beyond detection and utility, the weight-embedded baselines differ in their embedding and auditing cost. GaussMark's detector computes gradients of the suspect text's log-likelihood with respect to the perturbed parameters, which takes a forward and a partial backward pass for every audited text. \ourMethod needs one forward pass through the watermarked checkpoint to replay the routing and one through the clean checkpoint for the expected green mass $\mu_t$, followed by green-list lookups at the triggered positions, and no backward pass, which suits repeated, large-scale audits. Embedding procedures are similar for \ourMethod and for KGW-D in the logit-based form we compare against (Section~\ref{subsec:experiments_comparison}), since both distill the KGW distribution with LoRA. A sampling-based form of KGW-D would additionally require the generation of a watermarked training corpus of millions of tokens with a teacher~\cite{gu2024learnability}. However, embedding \ourMethod also trains and affects far fewer parameters (Section~\ref{subsec:payload_details}).

\textbf{Security of \ourMethodNoSpace.} The secrecy of \ourMethod rests on the seed, which hides the three targets among the candidate experts, and on targets that do not stand out in the weights. An adversary who knows the algorithm can still mask every candidate, but this comes at the cost of model utility (Section~\ref{subsec:broad_masking}), and the three weight-based signals we test do not single out the targets (Section~\ref{subsec:localization}). Fine-tuning on ordinary text still leaves 60.4 to 65.5\% of watermarked texts detected after 1,000 steps (Section~\ref{subsec:fine_tuning}). However, an adversary who also holds the clean checkpoint could find the target experts by comparing weights, so \ourMethod relies on the proprietary pre-watermarked weights staying private.

\textbf{Limitations.} Like other statistical text watermarks, \ourMethod requires sufficient textual evidence to accumulate a reliable detection signal. At 100 tokens, it detects about half of the watermarked texts (Appendix~\ref{subsec:seq_len}), and models with many routed experts accumulate evidence more slowly, because each token reaches a target expert less often. In practical auditing settings, both limitations can be partially mitigated by aggregating evidence across multiple texts attributed to the same suspect deployment. Defenders may alternatively target additional experts to increase trigger frequency, although Section~\ref{subsec:ablation_expert} shows that this comes at the cost of model utility. Among the attacks that leave the model untouched, paraphrasing reduces detection the most (Section~\ref{subsec:text_perturbations}), as is the case for token-level watermarks in general~\cite{rastogi2024paraphrase}. However, this requires the adversary to rewrite every text with a second model, while part of the signal can remain.



\section{Related Work}
\label{sec:related_work}

\textbf{Inference-Time LLM Watermarking.}
KGW~\cite{kirchenbauer2023watermark} introduces a keyed green-list bias during decoding and detects it through statistical testing. Subsequent methods improve distortion-free watermarking~\cite{kuditipudi2024robust}, adapt to next-token distributions~\cite{he2025theoretically}, optimize detectability--distortion trade-offs~\cite{gloaguen2026unifiedframeworkllmwatermarks}, and extend watermarking to diffusion language models~\cite{gloaguen2026diffusionwatermarking}. Post-generation approaches instead embed signatures into completed text~\cite{zhang2024remarkllm}. However, these methods rely on the owner's generation pipeline and cannot guarantee watermark persistence after model-weight theft~\cite{gu2024learnability}. In contrast, \ourMethod embeds the green-list bias directly into model weights.

\textbf{Weight-Embedded Watermarking.}
KGW-D~\cite{gu2024learnability} distills watermark patterns from a watermarked teacher into student weights, while related studies investigate watermarking open-source models~\cite{gloaguen2025opensource} and watermark transfer through fine-tuning~\cite{sander2024watermarking}. GaussMark~\cite{block2025gaussmark} embeds Gaussian weight perturbations and detects them using likelihood gradients. Although these approaches support text-based detection independently of the inference pipeline, they typically affect every token, creating utility and robustness trade-offs. Unlike these methods, \ourMethod confines watermark updates to a few secret experts, selectively influencing predictions through sparse routing.

\textbf{Watermarking MoE Models.}
WaterMoE~\cite{sun2026watermoe} introduces inference-time router biases to watermark generated text, but leaves model weights unchanged and therefore cannot survive weight theft. PathMark~\cite{gao2026pathmark} embeds a trigger-dependent routing signature into model weights, requiring queries or weight access for ownership verification rather than detecting provenance from ordinary generated text. In contrast, \ourMethod embeds the watermark into secret experts and reconstructs their routing from suspect text, enabling post-theft, text-only provenance verification. Unlike both approaches, its watermark also transfers to distilled dense students (Section~\ref{subsec:distillation}). Table~\ref{tab:related_work} summarizes these distinctions.

\section{Conclusion}
\label{sec:conclusion}
We present \ourMethodNoSpace, an architecture-aware watermarking framework that exploits sparse expert activation in MoE LLMs to enable post-theft, black-box text provenance. By embedding a statistical signal into selected experts and reconstructing their routing from suspect text, \ourMethod enables model-specific attribution without access to the adversary's deployment or weights. Across eight MoE architectures, \ourMethod achieves an average TPR of 92.5\% at 1\% FPR on 300-token texts, while reducing benchmark accuracy by only 0.4\% on average. Our security evaluation demonstrates resilience to quantization, fine-tuning, and distillation into dense models, while revealing the utility costs of expert masking and the limitations under paraphrasing. 
Overall, \ourMethod demonstrates how MoE's conditional computation can reconcile watermark detectability, utility preservation, and robustness in post-theft provenance verification.

\bibliographystyle{IEEEtran}
\bibliography{bibliography}

\appendices

\section{Effects of Sequence Length}
\label{subsec:seq_len}

\begin{figure}[!t]
  \centering
  \includegraphics{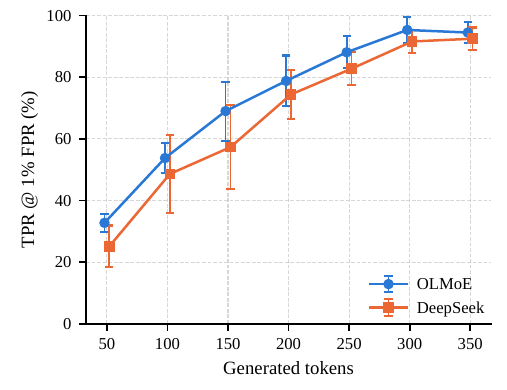}
  \caption{TPR at 1\% FPR (\%) against the number of generated tokens at the operating point $\delta^*$ on the test split. Points are means over the three seeds of each model, and bars show one standard deviation.}
  \label{fig:sequence_length}
\end{figure}

A defender often sees short texts, such as a post or a single answer, so detection has to work with as few tokens as possible. For this experiment, every text is generated to 350 tokens, and we score its prefixes in steps of 50 tokens, with thresholds recalibrated at each length on the tune-split null texts that reach it. Figure~\ref{fig:sequence_length} shows that detection builds up steadily with length. At 100 tokens, \ourMethod already detects 53.8\% of OLMoE's watermarked texts and 48.6\% of DeepSeek's, and at 50 tokens, 32.7 and 25.1\%.

\section{Settings Ablations}
\label{app:ablations}

\begin{figure*}[!t]
  \centering
  \includegraphics{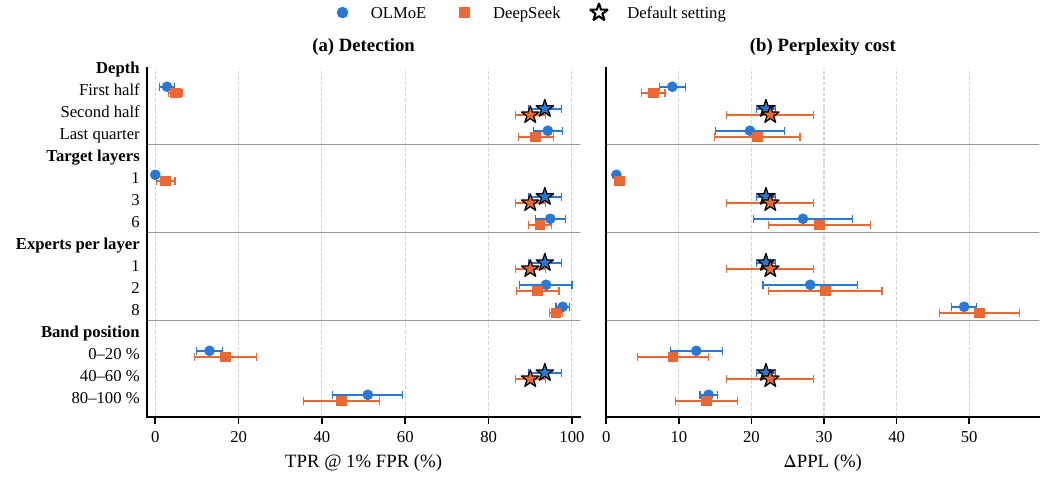}
  \caption{Settings ablations on the tune split. Each group of rows varies one setting from the default, with (a) the TPR at 1\% FPR (\%) and (b) the perplexity increase (\%) over the clean model of each level. Points are means over the three seeds of each model, and bars show one standard deviation. The star marks the default setting.}
  \label{fig:ablation}
\end{figure*}

Figure~\ref{fig:ablation} compares each level of each setting with the default by its TPR and its perplexity cost, and Table~\ref{tab:ablation} gives the values, together with the accuracy change of the benchmarked levels and the candidate experts an adversary without the seed has to mask at each level.


\begin{table*}[!t]
  \centering
  \footnotesize
  \caption{Settings ablations at $\delta^*$ on the tune split, as means over the three seeds of each model. For each axis and level, the TPR at 1\% FPR (\%), the mean $Z_{\mathrm{cal}}$, the trigger rate, the perplexity increase (\%) over the clean model, the change in mean benchmark accuracy (pp) for the benchmarked levels, and the candidate experts an adversary without the seed has to mask.}
  \label{tab:ablation}
  \begin{tabular}{llcccccc}
    \toprule
    Axis & Level & TPR (\%) & $Z_\mathrm{cal}$ & Trigger & $\Delta$PPL (\%) & $\Delta$Acc. (pp) & To mask \\
    \midrule
    \multicolumn{8}{l}{\textit{DeepSeek}} \\
    Depth & First half & 4.8 & 0.84 & 0.32 & +6.5 & -- & 169 \\
     & Second half (default) & 90.0 & 5.29 & 0.35 & +22.6 & $-$0.5 & 169 \\
     & Last quarter & 91.3 & 5.53 & 0.36 & +20.8 & -- & 91 \\
    Target layers & 1 & 2.5 & 0.51 & 0.11 & +1.8 & -- & 169 \\
     & 3 (default) & 90.0 & 5.29 & 0.35 & +22.6 & $-$0.5 & 169 \\
     & 6 & 92.3 & 5.97 & 0.58 & +29.4 & $-$0.9 & 169 \\
    Experts per layer & 1 (default) & 90.0 & 5.29 & 0.35 & +22.6 & $-$0.5 & 169 \\
     & 2 & 91.8 & 5.91 & 0.56 & +30.2 & -- & 169 \\
     & 8 & 96.2 & 8.86 & 0.92 & +51.4 & $-$1.5 & 169 \\
    Band position & 0--20\% & 16.8 & 1.52 & 0.22 & +9.2 & -- & 156 \\
     & 40--60\% (default) & 90.0 & 5.29 & 0.35 & +22.6 & $-$0.5 & 169 \\
     & 80--100\% & 44.7 & 2.71 & 0.33 & +13.8 & $-$0.4 & 169 \\
    \midrule
    \multicolumn{8}{l}{\textit{OLMoE}} \\
    Depth & First half & 2.8 & 0.97 & 0.39 & +9.1 & -- & 104 \\
     & Second half (default) & 93.5 & 5.76 & 0.37 & +22.0 & $-$0.2 & 104 \\
     & Last quarter & 94.2 & 6.24 & 0.38 & +19.8 & -- & 52 \\
    Target layers & 1 & 0.0 & 0.38 & 0.12 & +1.4 & -- & 104 \\
     & 3 (default) & 93.5 & 5.76 & 0.37 & +22.0 & $-$0.2 & 104 \\
     & 6 & 94.8 & 6.38 & 0.61 & +27.1 & $-$0.5 & 104 \\
    Experts per layer & 1 (default) & 93.5 & 5.76 & 0.37 & +22.0 & $-$0.2 & 104 \\
     & 2 & 93.8 & 6.15 & 0.58 & +28.1 & -- & 104 \\
     & 8 & 97.8 & 9.90 & 0.95 & +49.3 & $-$1.0 & 104 \\
    Band position & 0--20\% & 13.0 & 1.67 & 0.28 & +12.4 & -- & 96 \\
     & 40--60\% (default) & 93.5 & 5.76 & 0.37 & +22.0 & $-$0.2 & 104 \\
     & 80--100\% & 51.0 & 3.19 & 0.35 & +14.1 & \phantom{$-$}0.0 & 104 \\
    \bottomrule
  \end{tabular}
\end{table*}

\section{Worst-Case Threshold}
\label{app:statistics}

\begin{figure*}[!t]
  \centering
  \includegraphics{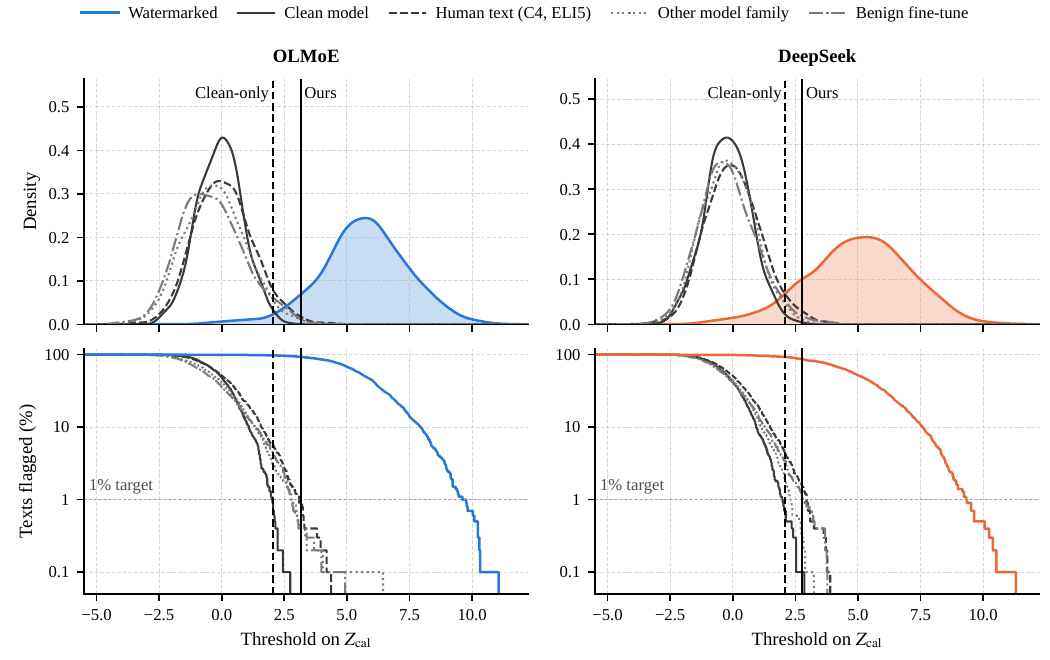}
  \caption{Calibrated scores $Z_{\mathrm{cal}}$ of watermarked text and of the four null sources at $\delta^*=3$ for one seed (A) of each model on the test split. The top panels show the density per text, and the bottom panels the share (\%) of each kind of text flagged as a function of the threshold (log scale), with the 1\% target dotted. Solid lines mark our worst-case threshold, set on the tune split, and dashed lines mark  a threshold set on the clean model's own text alone. On OLMoE, our threshold flags 0.9\% of human text, against 5.6\% under the clean-only threshold. On DeepSeek, it flags 1.5\% of human text, 0.3\% of the other family's text, 1.0\% of the benign fine-tune's text, and 0.1\% of the clean model's text, against 4.3, 1.9, 3.1, and 0.7\% under the clean-only threshold.}
  \label{fig:zcal_scores_appendix}
\end{figure*}

Figure~\ref{fig:zcal_scores_appendix} validates the choice of the worst-case threshold for one seed of each model. It shows the scores of watermarked text and of the four null sources, and how many texts of each kind a threshold flags.

\section{Per-Benchmark Accuracy}
\label{app:per_benchmark_accuracy}

Table~\ref{tab:benchmarks} breaks the mean accuracy change in Table~\ref{tab:main_results} down into its five benchmarks for the watermarked models, as mean and standard deviation over the three seeds, and for the benign fine-tune. Averaged over seeds, the watermarked models stay within 0.8\% of the clean model on MMLU, GSM8K, WinoGrande, and IFEval, while losing 1.0 to 3.5\% on HumanEval ($-$1.8\% on average). By contrast, the benign fine-tune loses at most 1.8\% on HumanEval ($-$0.6\% on average) but loses 2.5 to 4.3\% on GSM8K and 1.1 to 2.6\% on IFEval.

\begin{table*}[!t]
  \centering
  \footnotesize
  \setlength{\tabcolsep}{3.8pt}
  \caption{Per-benchmark accuracy change (pp) of \ourMethod at the operating point $\delta^*$ and of the benign fine-tune, relative to the clean model. The values of \ourMethod are mean $\pm$ SD over the three seeds of each model. The last two columns average the benchmarks of each model, and the last row averages the eight models.}
  \label{tab:benchmarks}
  \begin{tabular}{lcccccccccccc}
    \toprule
    & \multicolumn{2}{c}{MMLU} & \multicolumn{2}{c}{GSM8K} & \multicolumn{2}{c}{WinoGrande} & \multicolumn{2}{c}{IFEval} & \multicolumn{2}{c}{HumanEval} & \multicolumn{2}{c}{Average} \\
    \cmidrule(lr){2-3}\cmidrule(lr){4-5}\cmidrule(lr){6-7}\cmidrule(lr){8-9}\cmidrule(lr){10-11}\cmidrule(lr){12-13}
    Model & WoE & Benign & WoE & Benign & WoE & Benign & WoE & Benign & WoE & Benign & WoE & Benign \\
    \midrule
    DeepSeek & $-$0.4 $\pm$ 0.8 & $-$0.1 & \phantom{$+$}0.0 $\pm$ 0.6 & $-$3.7 & $-$0.5 $\pm$ 0.4 & $-$0.1 & $-$0.2 $\pm$ 0.6 & $-$1.8 & $-$1.6 $\pm$ 0.9 & $-$0.6 & $-$0.5 $\pm$ 0.2 & $-$1.3 \\
    GPT-OSS & $+$0.2 $\pm$ 0.3 & $-$0.5 & $-$0.4 $\pm$ 0.5 & $-$2.5 & $+$0.3 $\pm$ 0.6 & $+$0.5 & $+$0.2 $\pm$ 0.7 & $-$1.5 & $-$1.8 $\pm$ 1.2 & $-$1.2 & $-$0.3 $\pm$ 0.2 & $-$1.0 \\
    Hunyuan & $-$0.3 $\pm$ 0.5 & $+$0.2 & $+$0.3 $\pm$ 0.7 & $-$2.7 & $-$0.1 $\pm$ 0.3 & $-$0.2 & \phantom{$+$}0.0 $\pm$ 0.5 & $-$2.0 & $-$1.2 $\pm$ 1.2 & \phantom{$+$}0.0 & $-$0.3 $\pm$ 0.2 & $-$0.9 \\
    Mixtral & $-$0.6 $\pm$ 0.7 & $-$1.1 & $-$0.8 $\pm$ 1.0 & $-$3.9 & $-$0.2 $\pm$ 0.8 & $+$0.2 & $-$0.6 $\pm$ 0.7 & $-$2.6 & $-$3.5 $\pm$ 1.9 & $-$1.8 & $-$1.1 $\pm$ 0.6 & $-$1.8 \\
    Nemotron & $+$0.1 $\pm$ 0.4 & $-$0.7 & $-$0.1 $\pm$ 0.9 & $-$3.0 & $+$0.2 $\pm$ 0.5 & $-$0.4 & $-$0.4 $\pm$ 0.7 & $-$1.3 & $-$1.6 $\pm$ 1.5 & $-$0.6 & $-$0.4 $\pm$ 0.3 & $-$1.2 \\
    OLMoE & $-$0.1 $\pm$ 0.3 & $-$1.3 & $-$0.3 $\pm$ 0.8 & $-$4.3 & $+$0.7 $\pm$ 0.6 & $+$2.2 & $+$0.2 $\pm$ 0.6 & $-$2.2 & $-$1.4 $\pm$ 1.5 & \phantom{$+$}0.0 & $-$0.2 $\pm$ 0.2 & $-$1.1 \\
    Phi & $-$0.3 $\pm$ 0.6 & $-$0.9 & $+$0.2 $\pm$ 0.4 & $-$3.5 & $+$0.2 $\pm$ 0.7 & $+$0.7 & $-$0.1 $\pm$ 0.3 & $-$2.4 & $-$2.0 $\pm$ 1.5 & $-$1.2 & $-$0.4 $\pm$ 0.3 & $-$1.5 \\
    Qwen & \phantom{$+$}0.0 $\pm$ 0.2 & $-$0.4 & $-$0.6 $\pm$ 0.7 & $-$3.2 & $+$0.1 $\pm$ 0.5 & $+$0.3 & $+$0.1 $\pm$ 0.5 & $-$1.1 & $-$1.0 $\pm$ 0.9 & $+$0.6 & $-$0.3 $\pm$ 0.2 & $-$0.8 \\
    \midrule
    \textit{Average} & $-$0.2 & $-$0.6 & $-$0.2 & $-$3.4 & $+$0.1 & $+$0.4 & $-$0.1 & $-$1.9 & $-$1.8 & $-$0.6 & $-$0.4 & $-$1.2 \\
    \bottomrule
  \end{tabular}
\end{table*}

\end{document}